\documentclass[24pt]{article}
\usepackage{geometry}
\usepackage{xcolor}
\usepackage{a4}
\usepackage{alltt}
\usepackage{graphicx}
\usepackage{epsf}
\usepackage{amsmath}
\usepackage{amssymb}
\usepackage{cite}
\usepackage{hyperref}

\newcommand{\be}{\begin{equation}}
\newcommand{\ee}{\end{equation}}

\newcommand{\Rmnum}[1]{\expandafter\@slowromancap\romannumeral #1@}
\newcommand{\bea}{\begin{eqnarray}}
\newcommand{\eea}{\end{eqnarray}}
\begin{document}
\title{\bf Causal horizons in a bouncing universe}
\author{Pritha Bari, Kaushik Bhattacharya, Saikat Chakraborty 
\thanks{ E-mail:~ pribari, kaushikb, snilch@iitk.ac.in} 
\\ 
\normalsize
Department of Physics, Indian Institute of Technology,\\ 
\normalsize
Kanpur 208016, India}
\maketitle 
\begin{abstract}
As our understanding of the past in a bouncing universe is limited, it
becomes difficult to propose a cosmological model which can give some
understanding of the causal structure of the bouncing universe. In
this article we address the issue related to the particle horizon
problem in the bouncing universe models. It is shown that in many
models the particle horizon does not exist, and consequently the
horizon problem is trivially solved. In some cases a bouncing universe
can have a particle horizon and we specify the conditions for its
existence. In the absence of a particle horizon the Hubble surface
specifies the causal structure of a bouncing universe. We specify the
complex relationship between the Hubble surface and the particle
horizon when the particle horizon exists. The article also address the
issue related to the event horizon in a bouncing universe. A toy
example of a bouncing universe is first presented where we specify the
conditions which dictate the presence of a particle horizon. Next we
specify the causal structures of three widely used bouncing
models. The first case is related to quintom matter bounce model, the
second one is loop quantum cosmology based bounce model and lastly
$f(R)$ gravity induced bounce model.  We present a brief discussion on
the horizon problem in bouncing cosmologies.  We point out that the
causal structure of the various bounce models fit our general
theoretical predictions.
\end{abstract}
\section{Introduction}

The issue of causality is at the heart of relativistic physics. In
cosmology, where it is assumed that we live in an expanding universe
modelled on the Friedmann-Lemaitre-Robertson-Walker (FLRW) metric, the
causal nature of the universe is understood by the properties of the
particle horizon and the event horizon.  In Big-Bang cosmology it is
assumed that there is a finite age of the universe and intuitively one
can visualize that during this time light has travelled only a finite
region and consequently Big-Bang cosmology predicts a calculable
particle horizon for any observer in the present universe. On the
other hand if there was no Big-Bang to start with, as in non-singular
bouncing cosmological models\cite{Brandenberger:2016vhg,
  Battefeld:2014uga}, then it becomes very difficult to say whether a
particle horizon exists for any observer.  In this article we will
show that for non-singular bouncing cosmologies, where a contracting
phase precedes the expansion phase there may exist certain conditions
which dictate when a particle horizon will exist. Like the particle
horizon the event horizon plays an important role in gravitational
theories. In general thermodynamic behavior of a system can be linked
with the event horizon as is done in black hole physics.  Some authors
have even tried to associate thermodynamic properties of the universe
with the cosmic event horizon\cite{Gibbons:1977mu}.  In the present
article we will show that some of our examples of bouncing cosmologies
do have event horizons.

Except the particle and event horizons the Hubble radius also plays an
important role\cite{Lewis:2012yk, Harrison:1991dv} in determining the
causal structure of the universe.  The Hubble radius defines the
Hubble sphere and the surface of the Hubble sphere is called the
Hubble surface.  Many authors use Hubble horizon to describe the
boundary of the Hubble sphere although in this article we will not
associate the word horizon with the Hubble surface.  Later we will see
that in bouncing cosmologies the Hubble surface affects the causal
structure of spacetime in a very subtle way.  A general discussion
illuminating the relationship of the particle horizon and the Hubble
surface, in expanding spacetime, can be found in
Ref.\cite{Davis:2003ad}. The referred article addresses various
misconceptions related to the actual role of the Hubble surface.

Although the concepts of the particle horizon, event horizon and the
Hubble surface are commonly used and well discussed topics in Big-Bang
cosmology\cite{Ellis:2015wdi, Margalef-Bentabol:2012kwa,
  MargalefBentabol:2013bh} they are rarely discussed in the bouncing
universe paradigm where a contracting phase of the universe changes
course and starts to expand again giving rise to the expanding
universe we observe. The non-singular bouncing models are interesting
because they do not contain any
singularities\cite{Brandenberger:2016vhg, Battefeld:2014uga,
  Novello:2008ra, Martin:2004pm, Brandenberger:2012zb,
  Cai:2012va}. Once one includes a contracting universe, a universe
which does not have any initial time to start, the concept of the
particle horizon becomes much more involved. Some discussion on the
causality issue in bouncing cosmological scenario is included in
Ref.~\cite{Martin:2003bp}. In a bouncing universe the scale-factor of
the FLRW metric is not a power law function of time during the bounce
and consequently the difference between Hubble radius and the particle
horizon distance (if particle horizon exists) diverge maximally near
the bounce point. Before we end the our discussion on the causal
structure of purely bouncing models we want to briefly specify that
there are certain models of the early universe which employs a
non-singular bounce as well as cosmological inflation. In
Ref.~\cite{Martin:2003sf} the authors specify the necessity of
inflation after bounce and in Ref.~\cite{Cai:2014bea} the author
explicitly gives a model of matter bounce followed by inflation. We
will show in our article that matter bounce models, where the earliest
phase of the universe during the contraction phase was matter
dominated, lack particle horizons and in these models the whole of the
universe can be in causal contact. In matter bounce models one does
not require inflation to solve the horizon problem (the problem is
solved) but one may require inflation for other cosmological
purpose. In this article we will in general solely concentrate on the
causal structure of bouncing models which do not include inflation in
the beginning of the expansion phase.  The material in the article is
presented in the following way.  In the next section we describe the
preliminary theoretical tools which we will employ throughout the
article. In section \ref{pheh} we quantitatively define the concepts
of the horizons and the Hubble radius.  In section \ref{relc} we
specify a toy bounce model where the bouncing universe accommodates
two phase transitions during which the scale-factors of the metric
change.  In the subsequent section \ref{varb} we present three kind of
bouncing scenarios widely used by authors to design a cosmological
bounce. The first scenario deals with quintom matter bounce, the
second one with loop quantum cosmology induced bounce and the third
one is related to $f(R)$ gravity induced cosmological bounce. In
section \ref{hprob} we address the horizon problem in bouncing
cosmologies.  We discuss about the results obtained in this article in
section \ref{disc} and finally conclude the paper in the subsequent
section.
\section{Requirements of a cosmological bounce}

In this article we will use the homogeneous and isotropic FLRW
spacetime. We will particularly work with the spatially
flat FLRW metric and its form in the spherical polar coordinates is given by
\begin{eqnarray}
ds^2 = -dt^2 + a^2(t) \left[ dr^2 + r^2(d\theta^2 +
  \sin^2 \theta\,\,d\phi^2)\right]\,,
\label{sfrw}
\end{eqnarray}
where $r$ is comoving radial coordinate. We are using units where the
velocity of light $c=1$. Here $a(t)$ is the scale-factor for the FLRW
spacetime.  The Einstein equation in the cosmological setting can be
expressed in terms of the Hubble parameter $H\equiv \dot{a}/a$, where
a dot over a quantity specifies a time derivative of that quantity.
The relevant equations are
\begin{eqnarray}
H^2 &=&\frac{\kappa}{3}\rho\,,
\label{fried1}\\
\dot{H} &=&-\frac{\kappa}{2}(\rho + P)\,,
\label{fried2}
\end{eqnarray}
where $\rho$ is the energy density and $P$ is the pressure of matter
which pervades the universe. In the above equations $\kappa=8\pi G$
where $G=1/M_P^2$, $M_P=1.2 \times 10^{19}\,{\rm GeV}$, being the
Planck mass. We are using units where $\hbar=1$ where $\hbar$ is the
reduced Planck constant. We have assumed that the cosmological
constant term to be zero. The energy density and pressure are related
as
\begin{eqnarray}
P=\omega \rho\,,
\label{beos}
\end{eqnarray}
where the value of $\omega$ specifies a particular barotropic equation
of state.  During a contraction phase, the above equations always
predicts a spacetime singularity as $t \to 0$ (from the negative side)
if the matter content of the universe satisfies the strong energy
condition (SEC). One can evade the singularity, by introducing a
bouncing universe where the scale-factor remains finite as $t \to 0$,
by violating the null energy condition (NEC) in general relativity.
Till now we have specified the cosmological dynamics via the equations
of general relativity (GR), later in the article we will discuss
$f(R)$ gravity based cosmologies. The dynamical equations like
Eq.~(\ref{fried1}) and Eq.~(\ref{fried2}) will only get modified in
$f(R)$ gravity based cosmology. The other important point regarding
$f(R)$ theories is related to the energy conditions. The energy
conditions specify a bounce strictly in GR based models and we do not
give much importance to the energy conditions in modified gravity
theories as in these theories it becomes difficult to formally define
the energy conditions.

If one wants to avoid the Big-Bang singularity near $t=0$ then one can
model a universe where the scale-factor $a(t)$ never becomes zero at
$t=0$. In these models the universe contracts during the time $-\infty
< t \le 0$ and expands during $t \ge 0$ and the cosmological bounce
happens at $t=0$ when $a(t=0) \ne 0$ and $\dot{a}(t=0)=0$. The global
minima of the scale-factor is at the bounce point. The universe
expands after the bounce, implying $\dot{a}$ becomes positive and
increases after the bounce. This second condition implies that
$\ddot{a}(t=0)>0$. The above conditions can also be specified in terms
of the Hubble parameter as:
\begin{eqnarray}
\left. H \right|_{t=0}=0\,,\,\,\,\,\,
\left. \dot{H} \right|_{t=0} > 0\,.
\label{bcond}
\end{eqnarray}
The bounce conditions stated above and Eq.~(\ref{fried2}) immediately
shows that in the flat FLRW spacetime, during bounce
\begin{eqnarray}
\left[\rho + P \right]_{t=0} < 0\,.
\label{nec}
\end{eqnarray}
The above condition specifies that the NEC has to be violated at the
bounce point. The violation of the NEC, in the cosmological
perspective, require exotic matter as was inferred in
Ref.~\cite{Peter:2001fy}, later \cite{Rubakov:2014jja} also shows how
to conceive of NEC violating matter in the early universe. One may
even change the gravitational theory to accommodate a cosmological
bounce\cite{Paul:2014cxa, Bhattacharya:2015nda}. In this article we
will not go into the details of the complexities of theories which
generates a cosmological bounce, assuming satisfactory solutions of
these difficult issues exist in principle. We will concentrate on the
main issue of the article which is related to the causality question
in bouncing cosmologies.
\section{Particle Horizon, Event Horizon and Hubble Radius in bouncing
cosmological models}
\label{pheh}

In the context of, spatially flat, FLRW spacetime let us think of a
light ray which travels from a point $(t_i,R,\theta,\phi)$ to
$(t_0,0,\theta,\phi)$. Light travels along a null geodesic and in our
particular case we have assumed a null, radial geodesic which serves
our purpose. From the line-element we see that for such a null
geodesic
$$ds^2 = -dt^2 + a^2(t) dr^2= 0\,.$$
This above equation gives,
\begin{eqnarray}
\int_{t_i}^{t_0} \frac{dt}{a(t)} = R\,,
\label{tr}
\end{eqnarray}
which gives the comoving distance between the emitter and the
observer. In the above the subscript '$i$' refers to the time of
emission of the light signal from the source, and the subscript '$0$'
refers to the time of reception of the light signal by the
observer.

If the time $t_0$ is the present cosmological time when the observer
is observing the universe, the physical distance to the the emitter in
the observers frame will be $R_P(t_0) = a(t_0) R$.  The regions from
where light could reach the observer at $t=t_0$ forms the region which
may have any causal effect on the present condition of the universe
and regions outside this region can have no causal effect on the
present day universe. All the regions from which light has reached the
present observable universe is enclosed by the causal horizon for the
central observer. The causally connected region is enclosed by a
2-dimensional spacelike spherical surface whose radial physical
coordinate at $t=t_0$ is called $R_P$, and this spacelike surface is
called the particle horizon.  In the bouncing universe, $t_i \to
-\infty$, and we can write,
\begin{eqnarray}
R_P(t_0) \equiv a(t_0)\int_{-\infty}^{t_0} \frac{dt}{a(t)}\,. 
\label{pht}
\end{eqnarray}
We will use the above definition of the particle horizon throughout
this article.

If there exists a finite distance which light can travel in infinite
time in the future, emanating from a spatial point at some time, then
an event horizon can exist. The event horizon is defined by a spatial
two-dimensional spherical surface, whose radius is given by the
finite distance travelled by light in infinite time in the future.
The formal mathematical definition (of the radius) of the event
horizon is:
\begin{eqnarray}
R_E(t_0) \equiv a(t_0)\int_{t_0}^{\infty} \frac{dt}{a(t)}\,,
\label{eht}
\end{eqnarray}
where again $t_0$ is the present time of the observer. 

In bouncing cosmology it may happen that $R_P(t_0) \to \infty$ for
some finite $t_0$. In that case an observer at any cosmic time can get
information about the past universe without any bound and one says that the particle horizon does not exist. Similarly when one says the event horizon does not exist one means that
$R_E(t_0) \to \infty$ for some finite $t_0$. 

The Hubble radius is the radial coordinate of the boundary of the
Hubble sphere which is a closed two dimensional spatial surface at any
cosmological time. The Hubble radius is defined as:
\begin{eqnarray}
R_H(t_0) \equiv \frac{1}{|H(t_0)|}\,.
\label{hubs}
\end{eqnarray}
The center of the Hubble sphere is located at the observers
position. The Hubble sphere does not depend upon the past history or
the future of the universe. In the expression of $R_H$ we have
deliberately used the modulus of $H$ to accommodate a contracting
phase of the universe when $H<0$. 

From the definitions of the horizons one can deduce
\begin{eqnarray}
\dot{R}_P &=& 1 + H R_P\,,
\label{rpr}\\
\dot{R}_E &=& -1 + H R_E\,.
\label{rer}
\end{eqnarray}
In the Big-Bang paradigm it is seen that $R_P$ always increases
superluminally as $H R_P$ is positive definite. In the Big-Bang
paradigm the rate of increment of the particle horizon is more than
the expansion rate of the universe. The galaxies on the particle
horizon are receding with a speed $H R_P$ where as the particle
horizon is receding relatively faster and the size of the observable
universe increases with time. In the bouncing scenario more
interesting things can happen.

In the contracting phase of the universe $H<0$ and Eq.~(\ref{rpr})
shows that during this time
$$0 \le \dot{R}_P < 1\,,\,\,\,\,{\rm or}\,\,\,\,\,\,\dot{R}_P<0\,.$$
The particle horizon distance can increase with time when $H R_P > -1$
and the opposite can happen when $H R_P < -1$. If $HR_P=-1$ at an
instant of time, then at that instant $\dot{R}_P=0$ and consequently
bouncing cosmologies may have a minimum of the particle horizon
distance. In general
\begin{eqnarray}
  \ddot{R}_P
  = H +R_P(\dot{H}+H^2)\,.
\nonumber
\end{eqnarray}
When $HR_P=-1$ the double time derivative of the particle horizon is given by
\begin{eqnarray}
\ddot{R}_P = -\frac{\dot{H}}{H}\,,
\label{ddr}
\end{eqnarray}
which show that $HR_P=-1$ indeed predicts a minima of the particle
horizon distance, in the contracting phase when $H<0$, if
$\dot{H}>0$. The condition $\dot{H}>0$ holds true near bouncing time and in
this article we will see that the minima of the particle horizon
distance for most bouncing cosmologies (which admits a finite $R_P$)
appear in the contracting phase of the universe near the bounce time.
When the particle horizon distance increases with time, during the
contraction phase, one must have $|H| R_P < 1$ or
\begin{eqnarray}
0 \le \dot{R}_P < 1\,,\,\,\,\,\,\,{\rm if}\,\,R_P < R_H\,.
\label{rpin}  
\end{eqnarray}
Similarly, when the particle horizon distance decreases with time,
during the contraction phase, one must have
\begin{eqnarray}
\dot{R}_P <0\,,\,\,\,\,\,\,{\rm if}\,\,R_P > R_H\,.
\label{rpde}  
\end{eqnarray}
The above equations show that during the contraction phase of the
universe the particle horizon can increase subluminally or may
decrease at any rate. On the other hand if the condition $R_P = R_H$
holds for a certain time period then during that period $\dot{R}_P =
0$. 

From the above discussion one can predict some properties related to
the particle horizon and the Hubble radius. The points are as follows:
\begin{enumerate}
\item If the particle horizon distance tends to a constant,
  non-singular, value as $t_0 \to -\infty$ then the Hubble radius must
  be equal to the particle horizon distance as $t_0 \to -\infty$.

\item If $R_P(t_0) \to \infty$ as $t_0 \to -\infty$ then in the
  initial phase, or the full phase, of contraction one must have $R_P > R_H$.
\end{enumerate}
One can easily prove the above statements. If $R_P$ tends to a
constant as $t_0$ tends to large negative time then $\dot{R}_P \sim 0$
during a prolonged time period in the far past and consequently $R_P
\sim R_H$ as $t_0 \to -\infty$. The second statement can be proved by
noting the fact that if $R_P$ is maximum as $t_0 \to -\infty$ then for
later times the particle horizon distance can only decrease. When the
particle horizon distance decreases with time one must have $R_P >
R_H$. 

As we see that in bouncing cosmological models the particle horizon
distance may not always be increasing in the contracting phase a
natural question arises about the fate of wavelengths of cosmological
perturbations (or simply length scales). Can a physical wavelength
$\lambda_p$ once less than $R_P$ during contraction exceed the
particle horizon distance in the future? The answer to this question
was given in Ref.~\cite{Martin:2003bp}. Due to the importance of this
question we reproduce the result briefly in this section. Suppose the
physical wavelength of the mode at some time $t_c$ is given by
$\lambda_p(t_c)=2\pi a(t_c)/k$ where $k$ is the comoving wave
number. Let the particle horizon distance $R_P(t_c)$ be greater than
$\lambda_p(t_c)$ at time $t_c$. Once we know this we can now write for
any time $t_0>t_c$
\begin{eqnarray}
\frac{R_P(t_0)}{\lambda_p(t_0)} = \frac{k}{2\pi}\int_{-\infty}^{t_0}\,\frac{dt}{a(t)}
&=&\frac{k}{2\pi}\int_{-\infty}^{t_c}\,\frac{dt}{a(t)} + \frac{k}{2\pi}\int_{t_c}^{t_0}\,\frac{dt}{a(t)}\,,\nonumber\\
&=& \frac{R_P(t_c)}{\lambda_p(t_c)} + \frac{k}{2\pi}\int_{t_c}^{t_0}\,\frac{dt}{a(t)}\,.
\label{rpbl}
\end{eqnarray}
The first term on the right hand side of the second line is always greater than one by assumption and the term added to it is positive definite. Consequently once a physical length scale is inside the particle horizon it should always be inside the particle horizon. On the other hand a physical length scale once lesser than the Hubble radius may become super-Hubble during the contraction phase. The ratio
\begin{eqnarray}
\frac{R_H(t_0)}{\lambda_p(t_0)} = \frac{k}{2\pi|H(t_0)|a(t_0)}\,,
\label{rhbl}
\end{eqnarray}
can change from values greater than one to lesser than one, for some
values of $k$, when $R_H$ decreases during the contraction
process. These modes again re-enter the Hubble sphere near $t \to 0$
when $R_H \to \infty$. Consequently modes which are sub-Hubble can
become super-Hubble during the contraction phase of the universe but
such changes cannot happen if the initial modes are smaller then
$R_P$.

The way the event horizon grows with time is given in Eq.~(\ref{rer}).
In the contracting phase, where $H R_E$ is negative definite, it can
be easily seen that the event horizon will steeply decrease with time.
All the above points will become apparent in various bouncing models
in the later part of this article. It must be noted that the
properties of the horizons and Hubble radius discussed in this section
does not have any intrinsic connection with GR, the results discussed
in this section remains true even in scalar-tensor theories of
gravity.
\section{A toy model calculation of the horizons}
\label{relc}

In this section we present a toy calculation which shows how the
causal structures evolve in bouncing cosmologies based on GR. We use
the word ``toy model calculation'' because in this section we do not
specify the kind of ``matter'' which produces the cosmological
bounce. We primarily concentrate on the nature of the scale-factor
near bounce and simply assume that there must be some exotic matter
which induces the bouncing behavior. In the next sections we will
present various phenomenological models of cosmological bounce. The
discussion in this section will set the paradigm which will be used to
interpret later results.

\begin{figure}[h!]
\centering
\includegraphics[scale=.7]{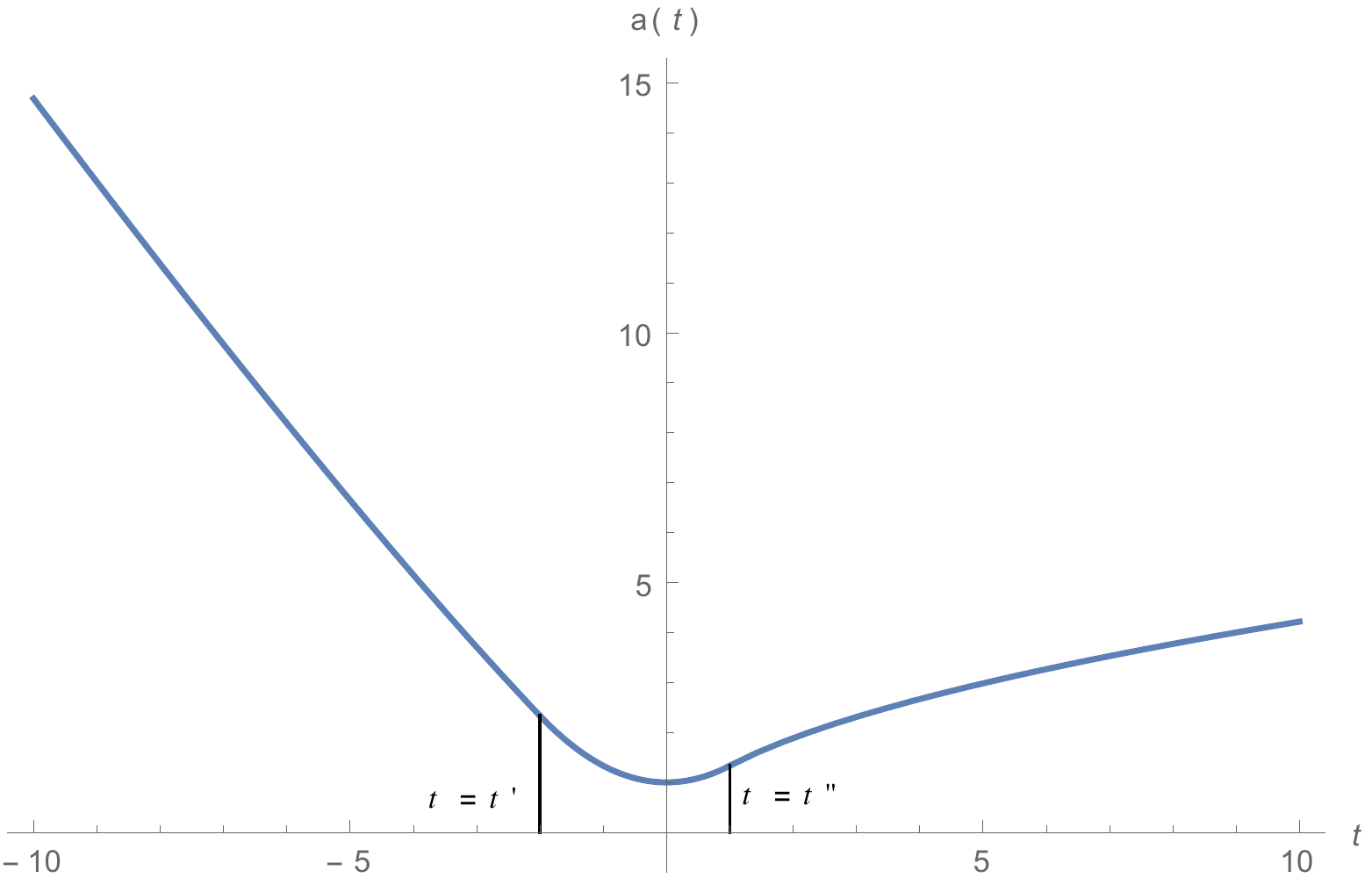}
\caption{The plot of the scale-factor for the bouncing case when
  $p=1$, $t^\prime=-2$ and $t^{\prime\prime}=1$.  The time period
  between the two black lines specify the bouncing phase. The other
  parameters used to plot the above graph are given in the discussion
  regarding the bounce for the $p=1$ case.}
\label{f:bexmp}
\end{figure}

In this section we assume at least three transformations (or phase
transitions in the matter sector) in the bouncing universe.  The first
transformation occurs during the contracting phase of the universe. We
assume that the contracting universe was dominated by some form of
matter which gave rise to a scale-factor whose functional form was
given by a power law. A power law scale-factor cannot lead to a
cosmological bounce and consequently we assume that at some time
$t^\prime < 0$ the nature of the matter content/geometry of the
universe changed. During $t^{\prime } < t < t^{\prime \prime}$ where
$t^{\prime \prime}>0$, the scale-factor of the universe changed from
the power law form and the new scale-factor accommodates a bounce at
$t=0$. Ultimately the universe comes out of the bouncing phase at
$t^{\prime \prime}$ and the scale-factor again transforms to a power
law function. This is a simple description of a cosmological bounce as
we know that the scale-factor of the bouncing universe must have
changed to a power law form after the bounce\footnote{In this article
  we do not consider the case where the bouncing universe culminates
  in an expanding inflationary universe.}. From our understanding of
the expanding phase of the universe and symmetry arguments it is
natural to think that some power law contraction phase may precede the
bouncing phase of the universe. The geometry of the universe changes
at $t^\prime$ and $t^{\prime\prime}$ and in our simplistic model the
change happens instantaneously. During such a phase transition we
assume that the junction conditions in GR are satisfied such that the
scale-factor and its time derivative remains continuous across the
junction\footnote{A nice discussion on the junction conditions in
  general relativity can be found in Ref.~\cite{poisson}.}.  The
change of the scale-factor in a typical bounce in our toy model is
shown in Fig.~\ref{f:bexmp}. The parameters specifying the bounce are
introduced later in this section.

We will assume that our model universe has the three following
scale-factors in the three different phases as: 
\begin{eqnarray}
a(t)=\left\{
\begin{array}{ll}
c_0 (-t)^m\,,\,\,\,\,\,\,\,\,t_i \le t < t^\prime\,,\\
a_0 + b_0 t^{2p}\,,\,\,\,\,\,\,\,\, t^{\prime } < t < t^{\prime \prime}\,, \\
d_0 t^n \,,\,\,\,\,\,\,\,\,t^{\prime \prime} < t \le \infty\,.\\
\end{array}
\right.
\label{atc}
\end{eqnarray}
In the above equation $t_i(<0)$ is assumed to be some initial time
which will ultimately tend to $-\infty$. The quantities $m$ and $n$
are positive real constants and $m$ is in general not equal to $n$.
The constant $p$ takes positive integer values.  Out of the three
constants we will assume that $0< n < 1$ as because in an expanding
flat FLRW model this constraint is generally followed. If the
scale-factor is given by a power law where $n$ is constrained in the
above way then the integral in Eq.~(\ref{eht}) diverges and the event
horizon does not exist.  The coefficients $c_0$, $a_0$, $b_0$ and
$d_0$ are also positive real constants out of which $a_0$ normalizes
the scale-factor at the bounce time.  The Hubble radius during the
three phases are:
\begin{eqnarray}
R_H(t_0) \equiv \frac{1}{|H(t_0)|}=\left\{
\begin{array}{ll}
\frac{|t_0|}{m}\,,\,\,\,\,\,\,\,\,t_i \le t_0 < t^\prime\,,\\
\left|\frac{a_0 + b_0 t_0^{2p}}{2b_0 p t_0^{2p-1}}\right|\,,\,\,\,\,\,\,\,\,
t^{\prime }
< t_0 < t^{\prime \prime}\, \\
\frac{t_0}{n} \,,\,\,\,\,\,\,\,\,t^{\prime \prime} < t_0 \le \infty\,\\
\end{array}
\right.
\end{eqnarray}
Because of the junction conditions, on the metric, at $t^\prime$ and
$t^{\prime\prime}$ one can easily verify that $R_H(t_0)$ changes
continuously at the junctions.  The particle horizon at any time
during expansion can be written as:
\begin{eqnarray}
R_P(t_0)= d_0 t_0^n \left [\int_{t_i}^{t'} \frac{dt}{c_0(-t)^m}+
  \int_{t'}^{t''} \frac{dt}{a_0 + b_0 t^{2p}} + \int_{t''}^{t_0}
  \frac{dt}{d_0t^n}\right]\,.\,\,\,\,\,\,\,\,(t^{\prime \prime} < t_0
\le \infty)
\label{rpd}
\end{eqnarray}
In the present case we have to specify the particle horizons in all
the three phases of development of the universe as the scale-factors
change during these phases. The particle horizon radius during the
power law contraction phase is given by
\begin{eqnarray}
  R_P(t_0) = \frac{1}{1-m}[(-t_0)^m (-t_i)^{1-m}+t_0]\,.\,\,\,\,\,\,\,\,
  (t_i \le t_0 < t^\prime)
\label{rpi}
\end{eqnarray}
Similarly, the particle horizon radius during the bouncing phase is given by
\begin{eqnarray}
R_P(t_0) = (a_0 +  b_0t_0^{2p})\left[\frac{1}{c_0(1-m)}\Big((-t_i)^{1-m}-(-t')^{1-m}
    \Big) +\int ^{t_0} _{t'} \frac{dt}{a_0 +  b_0t^{2p}}
  \right]\,.\,\,\,\,\,\,\,\, (t^{\prime } < t_0 < t^{\prime \prime})
\label{rpm}
\end{eqnarray}
The expression of the particle horizon distance during the expansion
phase is
\begin{eqnarray}
R_P(t_0)&=&d_0 t_0^n \left[\frac{1}{c_0(1-m)}\Big((-t_i)^{1-m}-(-t')^{1-m} \Big)
  +\int ^{t''} _{t'} \frac{dt}{a_0 +  b_0t^{2p}}\right.\nonumber\\
&+&\left.\frac{1}{d_0(1-n)}\Big(t_0^{1-n}-t''^{1-n}\Big)\right] \,.
\,\,\,\,\,\,\,\,(t^{\prime \prime} < t_0 \le \infty)
\label{rpf}
\end{eqnarray}
We will specify the integral containing the term $1/(a_0 + b_0
t^{2p})$ later, at present we concentrate on the junction
conditions. Applying the junction conditions at $t'$ we get
\begin{eqnarray}
  a_0=\left(\frac{2p}{m}-1\right)b_0 t'^{2p}\,,\,\,\,\,\,\,\,\,\,\,
  c_0 (-t')^m=\frac{a_0}{1-(m/2p)}\,.
\label{jptp}
\end{eqnarray}
Similarly applying the junction conditions at $t^{\prime\prime}$ one gets,
\begin{eqnarray}
a_0= \left(\frac{2p}{n}-1\right)b_0 t''^{2p}\,,\,\,\,\,\,\,\,\,\,\,
d_0 t''^n=\frac{a_0}{1-(n/2p)}\,.
\label{jctpp}
\end{eqnarray}
Comparing the above conditions one easily gets
\begin{equation}
\Big (\frac{t''}{t'}\Big)^{2p}  = \frac{\frac{2p}{m}-1}{\frac{2p}{n}-1}\,,
\label{tptpp}
\end{equation}
which sets a relationship between the matching times and the parameters
appearing in the scale-factors. Henceforth whenever we specify the
expression of the particle horizon and the Hubble radius we will
assume that the constants appearing in those expressions satisfy the
above junction conditions. 

The expressions of the particle horizon distances, as given in
Eqs.~(\ref{rpi}), (\ref{rpm}) and (\ref{rpf}), shows that $R_P(t_0)$
for all the phases is finite (as $t_i \to -\infty$) only when
$m>1$. Consequently, the existence of the particle horizon in the
realistic case depends upon the value of $m$.
\subsection{Nature of particle horizon}

In this subsection we will consider $m>1$ and assume $t_i \to -\infty$.
In this article we will present the results for two values of
$p$. In the first case $p=1$ and in the second case $p=2$ both of
which gives rise to a symmetric bouncing phase.
\subsubsection{The case where $p=1$}
\label{p1c}

When $p=1$ one can write the expressions for particle horizon distance as:
\begin{eqnarray}
R_P(t_0)=\left\{
\begin{array}{ll}
\frac{t_0}{1-m}\,,\,\,\,\,\,\,\,\,(-\infty \le t_0 < t^\prime)\\
(a_0 +  b_0t_0^2)\left[\frac{1}{c_0(m-1)}(-t')^{1-m} - \frac{1}{\sqrt{a_0
      b_0}} \arctan(\sqrt{\frac{b_0}{a_0}} t')+ \frac{1}{\sqrt{a_0
      b_0}} \arctan(\sqrt{\frac{b_0}{a_0}} t_0)\right]\,,
\,\,\,\,\,\,\,\,(t^{\prime } < t_0 < t^{\prime \prime}) \\  
d_0 t_0^n \left[\frac{1}{c_0(m-1)}(-t')^{1-m} - \frac{1}{\sqrt{a_0
      b_0}} \arctan\Big(\sqrt{\frac{b_0}{a_0}} t'\Big) + \frac{1}{\sqrt{a_0
      b_0}} \arctan\Big(\sqrt{\frac{b_0}{a_0}} t''\Big)\right.\\
\left.+\frac{1}{d_0(1-n)}\
\Big(t_0^{1-n}-t''^{1-n}\Big)\right]\,.\,\,\,\,\,\,\,\,\,(t^{\prime
  \prime} < t_0 \le \infty)
\end{array}
\right.
\end{eqnarray}
Finally we have to choose the constants appearing in the above
expressions judiciously such that the junction conditions are
satisfied. We take $a_0=1$ and $n=1/2$ assuming a radiation dominated
universe just after the bouncing phase. The time instants where the
scale-factors change are assumed to be $t'=-2$ and $t''=1$ in some
units. One can indeed express all time variables as multiples of a
fiducial time $t_f$, which can be a relevant microphysical one such as
some multiple of the Planck or GUT scale. One may choose $t_f=10^k
t_P$ where the $t_P \sim 10^{-43}$s is the Planck time and
$k>0$. Typically $k$ may be between 3 to 5. Once $t_f$ is specified
all the time labels as $t^\prime$ or $t^{\prime\prime}$ will actually
mean $t^\prime t_f$ or $t^{\prime\prime} t_f$. In this paper we will
always assume that all time intervals or time labels are actually
expressed in terms of the fiducial time unit $t_f$.
From the junction conditions one can now easily obtain 
$$m=\frac{8}{7}\,,\,\,b_0=\frac{1}{3}\,,\,\,c_0=\frac{7}{3}2^{-\frac{8}{7}}\,,\,\,
d_0=\frac{4}{3}\,.$$ Using these values we can write the particle
horizon radius at any phase of evolution of the universe. Particle
horizon distance at any time during power law contraction is simply
given by
\begin{eqnarray}
R_P(t_0)= -7t_0\,.
\end{eqnarray}
In this phase $t_0<0$. Particle horizon distance in the intermediate bouncing
phase comes out to be:
\begin{eqnarray}
R_P(t_0)= \Big(1 +  \frac{t_0^2}{3}\Big)\left[6 + \sqrt{3}
  \Big(\arctan\Big(\frac{t_0}{\sqrt{3}}\Big) + \arctan \Big(\frac{2}{\sqrt{3}}\Big)\Big)\right]\,.
\end{eqnarray}
Finally the particle horizon radius during the expanding phase of the
universe is:
\begin{eqnarray}
R_P(t_0)= \frac{4}{3} t_0^{\frac{1}{2}} \left[6 + \sqrt{3} \Big(\arctan{\Big(\frac{1}
    {\sqrt{3}}\Big)} +
  \arctan{\Big(\frac{2}{\sqrt{3}}\Big)}\Big)+\frac32(t_0^{1/2} -1)\right]\,.
\end{eqnarray}
\begin{figure}[t!]
\begin{minipage}[b]{0.5\linewidth}
\centering
\includegraphics[scale=.7]{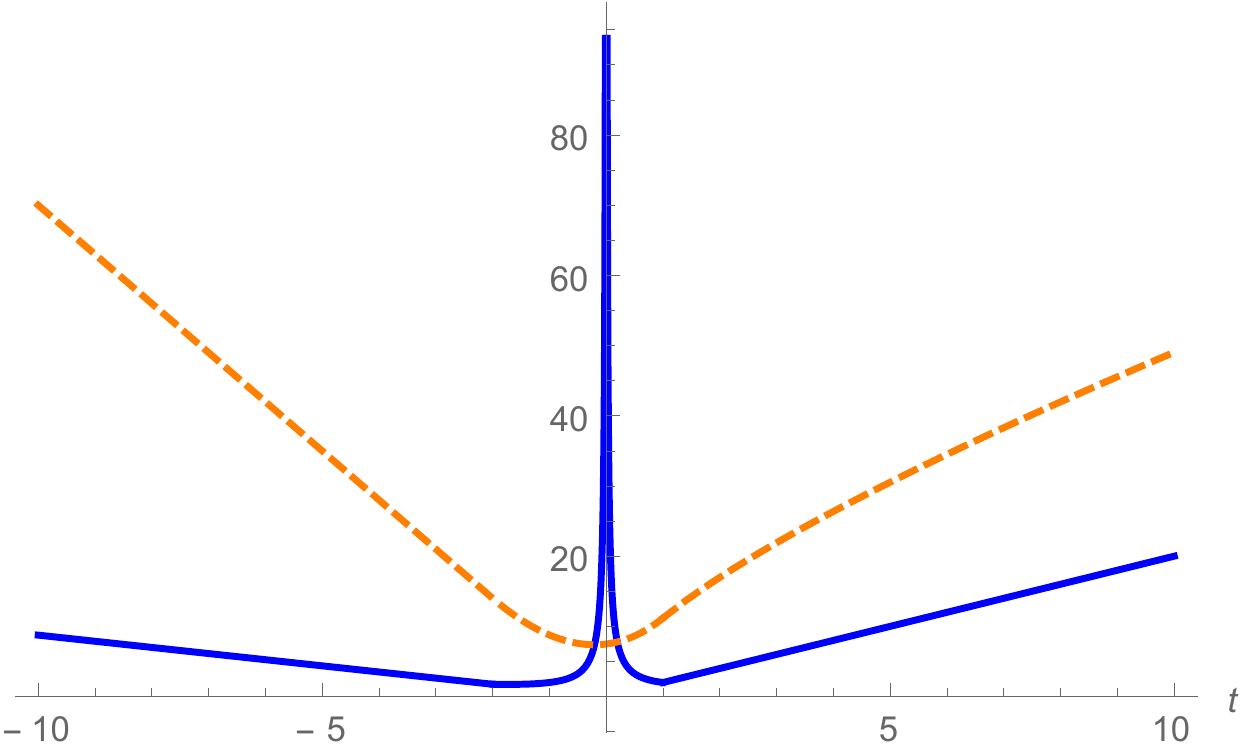}
\caption{Variation of particle horizon radius in orange dashed line
  and Hubble radius in blue through bounce for p=1.}
\label{f:p1}
\end{minipage}
\hspace{0.2cm}
\begin{minipage}[b]{0.5\linewidth}
\centering
\includegraphics[scale=.7]{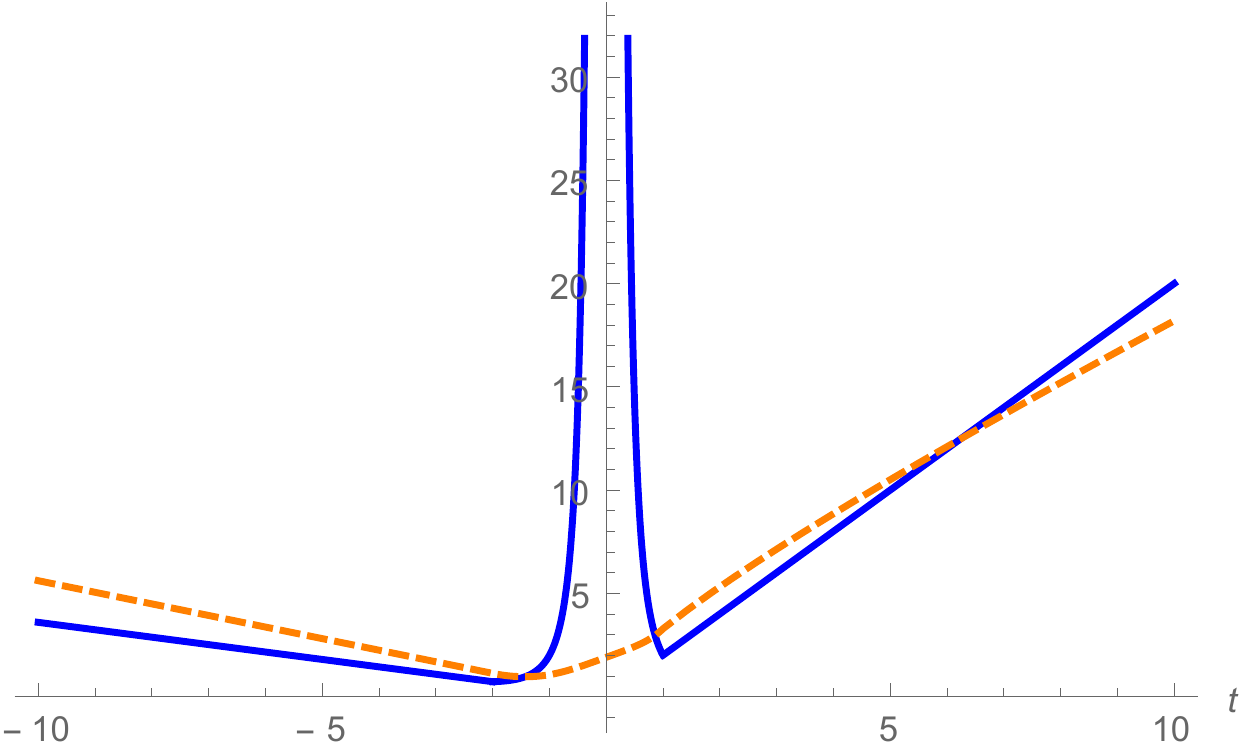}
\caption{Variation of particle horizon radius in orange dashed line
  and Hubble radius in blue through bounce for p=2.}
\label{f:p2}
\end{minipage}
\end{figure}
The particle horizon and the Hubble radius are plotted for the case
$p=1$ in Fig.~\ref{f:p1}. In the plot the Hubble radius is drawn in
blue and the particle horizon is shown by orange dashed curve.  The
plot shows that most of the time the Hubble sphere is causally
connected except very near to the bounce point where the Hubble radius
diverge. We have plotted the behavior of the particle horizon distance
and the Hubble radius near the bounce point and consequently the plots
do not convey the complete information about these distance scales
away from the bounce point. As one goes back in time the Hubble radius
increases and so does the particle horizon distance. The important
feature which comes out of the plot is that the particle horizon
decreases initially and then it attains a minimum value near the
bounce and increases in the expanding phase.
\subsubsection{The case where $p=2$}
\label{p2c}

In the present case the integral involving the term $1/(a_0 + b_0 t^{2p})$
yields\cite{grad}
$$\int\frac{dt}{a_0+b_0 t^4}=\frac{\alpha}{4 \sqrt{2}a_0}\left[\ln
  \Big(\frac{t^2 + \sqrt{2} \alpha t + \alpha^2}{t^2 - \sqrt{2} \alpha
    t + \alpha^2}\Big)+2 \arctan \frac{\sqrt{2}\alpha t}{\alpha^2 -
    t^2}\right]\,,$$ where $\alpha=(\frac{a_0}{b_0})^{\frac{1}{4}}$.
Using the above result we can write the particle horizon distance in
the three cases as:
\begin{eqnarray}
R_P(t_0)=\left\{
\begin{array}{ll}
\frac{t_0}{1-m}\,,\,\,\,\,\,\,\,\,(-\infty \le t < t^\prime)\\
(a_0 +  b_0 t_0^4)\left[\frac{1}{c_0(m-1)}(-t')^{1-m} +\frac{ \alpha}{
  4 \sqrt{2}a_0} \Big(\ln \Big[\frac{t'^2 - \sqrt{2} \alpha t' + \alpha^2}{
      t'^2 + \sqrt{2} \alpha t' + \alpha^2} \times \frac{
      t_0^2 + \sqrt{2} \alpha t + \alpha^2}{
      t_0^2 - \sqrt{2} \alpha t + \alpha^2}\Big]\right.\\
   \left. + 2 \arctan\frac{\sqrt{2} \alpha t_0}{\alpha^2 - t_0^2} - 
   2 \arctan\frac{\sqrt{2} \alpha t'}{\alpha^2 - t'^2}\Big)\right]\,,\,\,\,\,\,\,\,\,
(t^{\prime } < t_0 < t^{\prime \prime})\\
d_0 t_0^n \left[\frac{1}{c_0(m-1)}(-t')^{1-m}  +\frac{ \alpha}{
  4 \sqrt{2}a_0} \Big(\ln\Big[\frac{t'^2 - \sqrt{2} \alpha t' + \alpha^2}{
      t'^2 + \sqrt{2} \alpha t' + \alpha^2} \times \frac{
      t''^2 + \sqrt{2} \alpha t'' + \alpha^2}{
      t''^2 - \sqrt{2} \alpha t'' + \alpha^2}\Big]\right. \\
\left. + 2 \arctan\frac{\sqrt{2} \alpha t''}{\alpha^2 - t''^2} - 
2 \arctan\frac{\sqrt{2} \alpha t'}{\alpha^2 - t'^2}\Big)+
\frac{1}{d_0(1-n)}(t_0^{1-n}-t''^{1-n})
\right]\,.\,\,\,\,\,\,\,\,(t^{\prime \prime} < t_0 \le \infty)\\
\end{array}
\right.
\end{eqnarray}
In the present we set $a_0=1$ and $n=1/2$ as done in the previous
case.  The time instants where the scale-factors change are assumed to
be the same as in the previous case. The junction conditions now
predict
$$m=\frac{64}{23}\,,\,\,b_0=\frac{1}{7}\,,\,\,c_0=\frac{23}{7}2^{-\frac{64}{23}}\,,\,\,
d_0=\frac{8}{7}\,.$$ The resulting particle horizon distance is
plotted in orange dashed curve in Fig.~\ref{f:p2}. The Hubble radius
at each instant is plotted in blue curve. The present plot is
qualitatively same as the one in Fig.~\ref{f:p1}. The only difference
between them is that for the case $p=2$ the radiation dominated
universe can have a certain region where the Hubble radius exceeds the
particle horizon distance. Both the curves show that the particle
horizon follows a smooth curve which has a minima near the bounce
point.

In both the above cases we observe that during the contraction phase
the Hubble surface lies within the particle horizon. From our
discussion in section \ref{pheh} one can infer that in such cases the
particle horizon size must decrease with time. 
\subsection{Effect of the various parameter choices on the toy bounce model}

The generic features of the particle horizon were discussed in section
\ref{pheh}. Except the generic features many properties of the
cosmological system may depend on parameter choices. We will like to
end our discussion on the toy model by presenting a more general
discussion on the dependence of the system on various parameters
appearing in the model.

From the expression of the scale-factors in the various phases, as
given in Eq.~(\ref{atc}), we see that our model has in total nine
parameters as: $n$, $p$, $m$, $d_0$, $a_0$, $b_0$, $c_0$, $t^\prime$
and $t^{\prime \prime}$.  Most of these parameters have some
conditions as $m$, $n$ and $p$ are positive real constants and
$t^\prime<0$ and $t^{\prime \prime}>0$. The definition of the various
cosmological phases justifies the above constrains. More over as
because the scale-factor cannot be negative we must have $d_0$ and
$c_0$ to be positive and real constants. For a proper bouncing
behavior near $t=0$ one must also require $a_0$ and $b_0$ to be
positive real constants.  

Out of the nine parameters discussed above only five can be
independently chosen (all of which satisfies the conditions discussed
in the last paragraph) because of the relations appearing in
Eq.~(\ref{jptp}) and Eq.~(\ref{jctpp}). In the results presented we
have independently chosen $p$, $a_0$, $n$, $t^\prime$ and $t^{\prime
  \prime}$. The non existence of event horizons in our case was a
result of choosing $0<n<1$. If we had chosen $n>1$ our models will
also have event horizons. We have chosen $a_0=1$ for both the cases as
this choice normalizes the scale-factor at the bounce point. We will
like to use this normalization as no new physics emerges by changing
this normalization.  Except this choice we have also chosen $p$ to be
positive integers as 1 and 2. While in principle $p$ can take
non-integer values, we will show below that taking $p$ to be a
non-integer value does not alter the physics of the model.

The important thing to be discussed in this section is that the particle
horizon exists only when $m>1$.  In this subsection we will see that
the choice of values of $t^\prime$ and $t^{\prime \prime}$ does have
very interesting consequences as far as the existence of the horizon
is concerned. If we call the ratio $t^{\prime \prime}/t^\prime = t_r$
then Eq.~(\ref{tptpp}) gives us
$$t_r^{2p} = \left(\frac{n}{m}\right)\frac{2p-m}{2p-n}\,,$$
then it immediately gives some more constrains on the parameters, as
$2p>m$ and $2p>n$ simultaneously or $2p<m$ and $2p<n$
simultaneously. More over the condition for the existence of particle
horizon translates to
\begin{eqnarray}
t_r^{2p} < n\left(\frac{2p-1}{2p-n}\right)\,.
\label{trc}
\end{eqnarray}
The above relation shows that $t_r$, $p$ and $n$ must satisfy an
inequality if a particle horizon exists in the toy bouncing model. The
parameter choices in the previous subsections all satisfy the above
conditions.
\begin{figure}[t!]
\centering
\includegraphics[scale=.8]{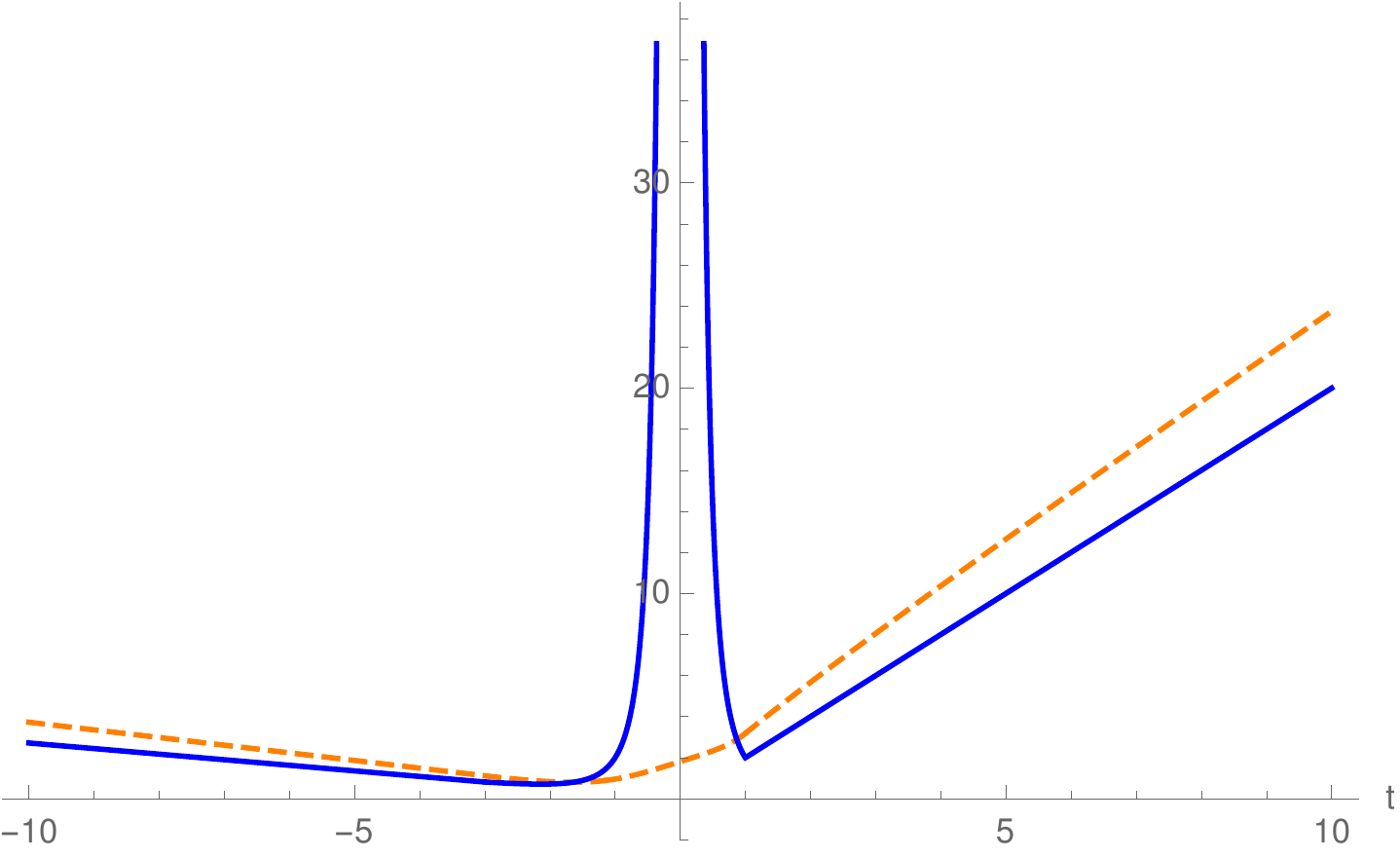}
\caption{The case where $p=2$ but $t^\prime=-3$ and
  $t^{\prime\prime}=1$. In this case the Hubble radius remains smaller
that $R_P$ during the expansion phase away from the bouncing point.}
\label{f:hdiff}
\end{figure}

During the expansion phase in standard (Big-bang) model of cosmology
the Hubble radius remains proportional to the particle horizon
distance when the scale-factor is a power law function of time. In
general the particle horizon distance always remains greater than or
equal to the Hubble radius when the power law index (of the
scale-factor) is a fraction, in Big-bang cosmologys. For the radiation
dominated early universe one gets $R_p=R_H$. For power law expansion
phases one never encounters a situation when $R_H$ becomes greater
than $R_P$ in standard cosmology. If in the bouncing models of
cosmology one demands that the radiation dominated phase just after
bounce develops similarly to the standard cosmological models then the
$p=2$ case as shown in our earlier subsection has to be rejected. If
one does not reject the cases as studied above then one may obtain
interesting results which may have observational signatures. We
discuss this issue in the section devoted to the horizon problem in
bouncing cosmological models.

One can choose parameters in the model such that $R_H$ never becomes
bigger than $R_P$ in the expansion phase of the toy models. The result
of one of such parameter choices is shown in Fig.~\ref{f:hdiff}.  In
this case all the parameters except the value of $t^\prime$ remains
the same as in the case of sub-subsection \ref{p2c}.  Till no new
evidence of bouncing cosmologies are observed from the cosmic
microwave background radiation (CMBR) data one may be tempted to avoid
bouncing models where at any time $t_0$ during the expansion phase of
the universe one gets $R_P(t_0)=R_H(t_0)$ and
$\dot{R}_P(t_0)=\dot{R}_H(t_0)$. In such cases the bouncing models
will be further constrained.
\section{The horizons in various different bouncing models}
\label{varb}

After the general discussion on the toy bounce model we will discuss
some specific models of cosmological bounce in this section. The first
subsection deals with bounce influenced by quintom matter. The second
subsection deals with loop quantum cosmology influenced cosmological
bounce. In the third subsection we do present a particular bouncing
solution in $f(R)$ cosmology. We do not claim that our examples are
exhaustive in character but the results presented in this section does
indeed reveal the causal structure of some important cosmological
bounce models. Unlike the toy bounce model we do not specify various
phase transitions of the universe, undergoing cosmological bounce, in
the cases discussed. For simplicity we have not separated the exact phases and
scale-factors in the examples given in this section although the
connection with the results of the toy-model will become apparent and
will be pointed out at various positions.
\subsection{The causal structure in quintom bounce model}
In the quintom bounce model the ``matter'' in the universe transits
from a $\omega > -1$ phase during contraction to the bouncing phase
where $\omega < -1$ and NEC is violated. In the initial expanding
phase after bounce the equation of state again changes from $\omega <
-1$ to $\omega > -1$ and the universe enters a hot Big Bang era after
the bounce. There are many ways one can build quintom models, perhaps
the simplest model consists of two scalar fields which produces the
quintom nature. Out of the two scalar fields one is similar in
character to the quintessence field and the other one is similar to
the phantom field \cite{Cai:2009zp, cai2, Feng:2004ad}.
\begin{figure}[t!]
\centering
\includegraphics[scale=.8]{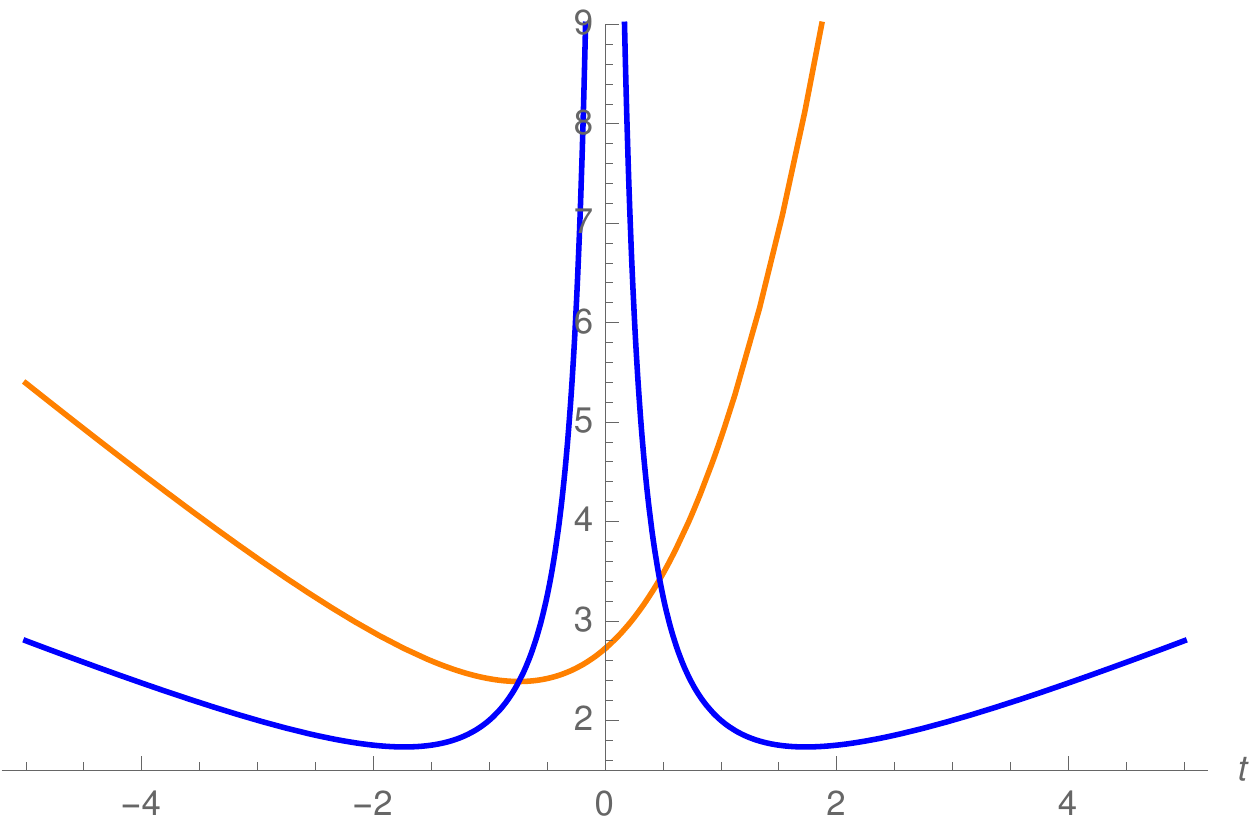}
\caption{Variation of the particle horizon distance (in orange) and
  Hubble radius (in blue) in quintom bounce model.}
\label{f:quint}
\end{figure}

In the quintom bounce model one can use an equation of state 
described by
\begin{eqnarray}
\omega = -r - \frac{s}{t^2}\,,
\label{qui}
\end{eqnarray}
where $t$ is cosmological time.  Here $r$ is a dimensionless constant
where as $s$ is a dimensional constant \cite{Cai:2009zp}. In general
$0<r<1$ and $s>0$ such that away from bounce one has $\omega>-1$ and
near the bouncing point, $t=0$, $\omega \ll -1$. Solving the Friedmann
equations in GR using the above equation of state one gets:
\begin{eqnarray} 
a(t)=\left[t^2 + \frac{s}{1-r}\right]^{\frac{1}{3(1-r)}}\,,
\label{atquin}
\end{eqnarray}
which shows the signature of a non-singular bounce at $t=0$ when the
equation of state diverges. If we compare the form of the scale-factor
in the quintom bounce case with the bouncing scale-factor in the toy
model, given by the middle expression in Eq.~(\ref{atc}), we see that
near the bounce point the above scale-factor is similar in form to the
one given in the toy model with $p=1$.  From Eq.~(\ref{atquin})
we see that as $t \to -\infty$ the scale-factor behaves as $a(t) \sim
(-t)^m$ where
$$m=\frac{2}{3(1-r)}\,,$$ which shows that $m>1$ for $r>1/3$. We know
from the toy example that in such a case the universe can admit a particle horizon:
\begin{equation} 
\begin{split}
R_P(t_0) & = \Big (t_0^2 +\frac{s}{1-r}\Big)^{\frac{1}{3(1-r)}} \int_{- \infty}^{t_0}
\frac{dt}{a(t)} \\
 & = \Big(t_0^2 +\frac{s}{1-r}\Big)^{\frac{1}{3(1-r)}} \int_{- \infty}^{t_0} \Big(t^2
+\frac{s}{1-r}\Big)^{-\frac{1}{3(1-r)}}dt
\end{split}
\end{equation}
Consequently the quintom models may or may not accommodate a particle
horizon.  The criterion for existence of the particle horizon depends
solely upon the numerical value of the dimensionless constant $r$. In
this case the expression of the Hubble parameter is
\begin{eqnarray}
H(t)=\frac23\left[\frac{t}{(1-r)t^2 + s}\right]\,,
\label{hubquin}
\end{eqnarray}
which helps us to specify the nature of the Hubble surface through out
the contraction phase of the universe. In Fig.~\ref{f:quint} we show
the nature of variation of the particle horizon distance and the
Hubble radius during the bounce. In this case we have used $s=1$ and
$r=2/3$. The causal structure of the quintom matter influenced bounce
show that the particle horizon distance, in general, is always greater
than the Hubble radius except at a narrow region near the bouncing
point where the condition $R_P=R_H$ is satisfied. The particle horizon
radius in Fig.~\ref{f:quint} does not say the general nature of all
quintom matter influenced cosmological bounces, as there is the other
case where $r<1/3$ when the particle horizons does not exist. In those
cases the Hubble radius specifies the causal structure of the quintom
bounce models. From Fig.~\ref{f:quint} we see clearly that the minimum
of the particle horizon distance happens when $R_P=R_H$ during the
contraction phase, as discussed in section \ref{pheh}.
\subsection{The evolution of the Hubble radius in loop quantum cosmology }
Loop quantum gravity (LQG) is a background independent
non-perturbative quantized theory of gravity. Loop quantum cosmology
(LQC) tries to understand the cosmological evolution, near the initial
singularity, of our universe in the light of some simple ideas coming
from LQG. In loop quantum cosmology (LQC), spacetime is quantized by
using the holonomies of SU$(2)$ group.  From the Hamiltonian
constraint one can derive the effective equations of motion, guiding
the dynamics of the universe, that include the leading order quantum
corrections to the classical equations of general relativity. In LQC
the evolution of the universe takes place in the following way. The
early time singularity is avoided by quantum effects and the universe
is guided by quantum laws during this phase. The early quantum epoch
takes into account the discreteness of spacetime near singularity, and
it was shown that the initial singularity can be avoided to achieve
both inflationary and bouncing cosmology (\cite{Ashtekar:2011ni},
\cite{Oikonomou:2014jua}, \cite{Singh:2006im}). In LQC the singularity
is avoided by a cosmological bounce guided by quantum gravity laws.
Next there appears a semiclassical epoch when some of the quantum
correlations transform into classical signals. Ultimately the
universe transforms into a classical phase.

For semi-classical states the quantum gravity effects are well
approximated by a set of effective equations. For the flat FLRW
universe the effective Friedmann equations which captures the quantum
effects are\cite{Cai:2014zga, deHaro:2012cj},
\begin{eqnarray}
H^2 &=& \frac{\kappa \rho}{3}\left(1-\frac{\rho}{\rho_c}\right)\,,
\label{h2lqc}\\
\dot{H} &=& -\frac{\kappa}{2}(\rho + P)\left(1-\frac{2\rho}{\rho_c}\right)\,,
\label{hdlqc}
\end{eqnarray}
where the critical energy density $\rho_c \sim M_P^4$. In the above
effective equations the other terms have their conventional meaning
and $H$ is still defined as the ratio of the time derivative of the
scale-factor to the value of the scale-factor. When the energy density
of the universe reaches $\rho_c$ the universe undergoes a cosmological
bounce. The matter variables $\rho,\,P$ can specify conventional
matter as dust or radiation. In both \cite{Cai:2014zga, Haro:2017mir}
the authors talk about matter bounce, where in the long past of the
contraction era the universe was filled with pressure-less matter and
then there was a phase transition at some negative time (much before
bounce) when the universe entered the ekpyrotic phase. The ekpyrotic
phase is obtained by using some specific form of scalar field
potential during the contraction phase.
\begin{figure}[t!]
\centering
\includegraphics[scale=.8]{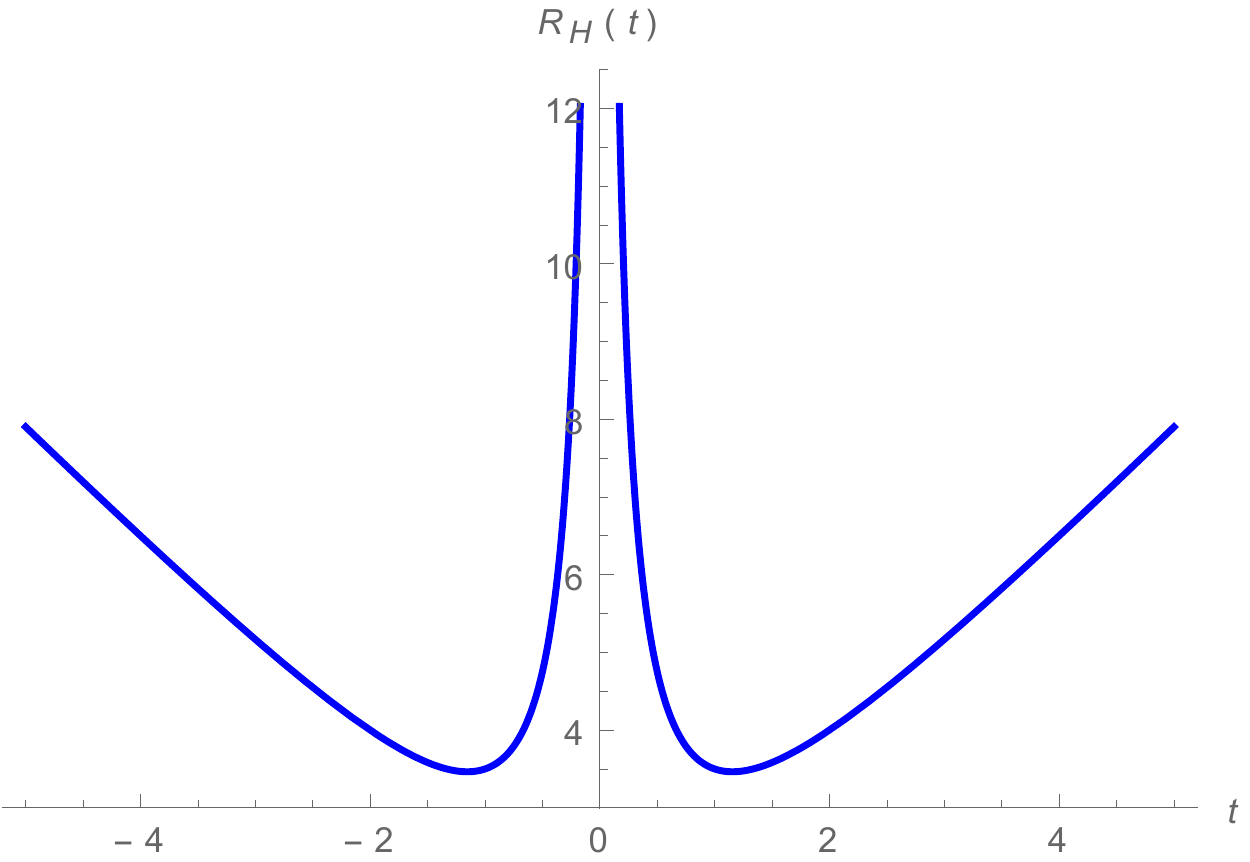}
\caption{Variation of the Hubble radius near the bounce point in LQC
  motivated matter bounce.}
\label{f:lqc}
\end{figure}
In \cite{deHaro:2012cj} we notice that as one moves away from the
bounce in the past the scale-factor becomes proportional to
$(-t)^{2/3}$ when $\omega=0$. Both in the ekpyrotic models and the
model in \cite{deHaro:2012cj} we see far away in the past we have the
scale-factor behaving as $(-t)^{2/3}$ and consequently in these LQC
motivated matter bounce models we will not have any particle horizon,
a point established in our analysis of the toy bounce model
study. From Ref.~\cite{deHaro:2012cj} we see that the scale-factor of
the universe in the immediate vicinity of $t=0$, when $\omega=0$, is given by
\begin{eqnarray}
a(t)=\left(1+\frac34 \kappa \rho_c t^2\right)^{1/3}\,,
\label{lqca}
\end{eqnarray}
from which we can specify the Hubble radius near about the bounce
point. Near $t=0$ we see that the above scale-factor becomes similar
in form to the scale-factor, given by the middle expression on the
right hand side in Eq.~(\ref{atc}), in the toy model bounce with
$p=1$. In Fig.~\ref{f:lqc} we plot the effective nature of variation
of the Hubble radius near the bounce point. In this plot we have used
$\kappa\rho_c=1$, which becomes natural if one sets $8\pi M_P^2=1$. In
absence of any particle horizon the Hubble radius sets the causal
structure of these cosmologies.
\subsection{The causal structure of $f(R)$ theory induced bounces}
\label{tbm}

In $f(R)$ gravity, most of the basics we have discussed in sections 2,3  remain the same except the dynamical equations of cosmology as given in Eq.~\ref{fried1}  and Eq.~\ref{fried2}. The new equations replacing them are:
\begin{eqnarray}
H^2 &=&\frac{\kappa}{3}(\rho + \rho_{\rm curv})\,,
\label{fried1r}\\
\dot{H} &=&-\frac{\kappa}{f^\prime(R)}(\rho + P + \rho_{\rm curv} + P_{\rm curv})\,,
\label{fried2r}
\end{eqnarray}
where a prime over a quantity designates a derivative with respect to
the Ricci scalar $R$. The curvature dependent energy-density and
pressure are
\begin{eqnarray}
  \rho_{\rm curv} &=& \frac{Rf^\prime -f}{2\kappa} - \frac{3Hf^{\prime\prime}\dot{R}}{\kappa}\,,
  \label{rhofr}\\
  P_{\rm curv} &=& \frac{\dot{R}^2 f^{\prime\prime\prime} + 2H\dot{R}f^{\prime\prime}
  +\ddot{R}f^{\prime\prime}}{\kappa} -  \frac{Rf^\prime -f}{2\kappa}\,.   
\label{pfr}
\end{eqnarray}
The bouncing conditions remain the as in GR. Simple forms of $f(R)$
may accommodate cosmological bouncing solutions. The simplest of them
may be when $f(R)=\lambda + R + \alpha R^2$ where $\alpha<0$ for a
bounce. The cosmological bounce in this model where $\lambda=0$ was
studied earlier \cite{ruzmaikina1970sov}. Later a thorough analysis of
such a bounce was also presented in \cite{Paul:2014cxa}. Bouncing in
quadratic gravity with $\lambda \ne 0$ has been studied in
\cite{Bamba:2013fha}. In the last case the authors note that a solution of the bouncing problem in quadratic gravity with non-zero $\lambda$ can be written as
\begin{eqnarray}
a(t)= a_0 e^{A t^2}\,,
\label{atfr}
\end{eqnarray}
where $a_0$ and $A$ are constants. The quadratic gravity model which
accommodates a cosmological bounce is in general gravitationally
unstable. The instability arises because of the fact that
$f^{\prime\prime}(R)<0$ for $\alpha<0$. There can be another
instability, when $f^\prime(R)<0$ in some domain of the Ricci
scalar. One can suitably choose the parameters such that
$f^\prime(R)>0$ during the bouncing regime. The above form of
scale-factor can also give a cosmological bounce when
$f(R)=\frac{1}{\beta}e^{\beta R}$ where $\beta>0$ is some constant
parameter of the theory. This theory does not have the above mentioned
instabilities. There can be many other forms of $f(R)$ which admits
Eq.~(\ref{atfr}) as a solution. In this section we will specify the
causal structure of an universe, where the underlying theory of
gravity is $f(R)$ gravity and the scale-factor of the universe is as
given in Eq.~(\ref{atfr}). In this case also we notice that the
scale-factor used becomes similar in form to the middle expression on
the right hand side in Eq.~(\ref{atc}) in the toy model bounce with
$p=1$ \footnote{It must be noted that in the present case one cannot
  match the scale-factors at various time instants as was done in the
  toy bounce model. The matching of scale-factors should be more
  involved in the present case as the junction conditions in general
  change in $f(R)$ gravity when compared with the same conditions in
  GR.}.  The functional form of the scale-factor show that the
bounce is symmetrical in time.  The particle horizon is given as
\begin{eqnarray}
R_P(t_0) =  e^{A t_0^2} \int_{-\infty}^{t_0} e^{-A t^2} dt\,,
\end{eqnarray}
assuming the scale-factor remained the same till $t\to -\infty$. The
integral on the right hand side can be evaluated by using the properties of the
Gaussian integrals. After doing the integral one obtains
\begin{eqnarray}
\int_{-\infty}^{t_0} e^{-A t^2} dt=\sqrt{\frac{\pi}{A}}\left[1
  -\frac{1}{2}\,\, {\rm erfc}(\sqrt{A}\, t_0)\right]\,,
\label{phint1}
\end{eqnarray}
where ${\rm erfc}(\sqrt{A}\, t_0)$ is the complimentary error
function.  To see how the above integral behaves in the extreme cases
where $t \to \pm\infty$ one requires the following properties of the
complementary error function,
$$\lim_{x \to -\infty} {\rm erfc}(x)=2\,,\,\,\,\, \lim_{x \to
  \infty} {\rm erfc}(x)=0\,.$$ It must be noted that the above
limits saturates near $x=0$, this information
will help us to figure out the behavior of the particle horizon at
the extreme limits. With all these information we can now write the
expression for particle horizon as:
\begin{eqnarray}
R_P(t_0) =  e^{A t_0^2} \sqrt{\frac{\pi}{A}}\left[1
-\frac{1}{2}\,\,
{\rm erfc}(\sqrt{A}\, t_0)\right]\,.
\label{case1}
\end{eqnarray}
The limits and their properties of the complimentary error function
listed above shows that the value of the particle horizon remains
finite at both the extremities of the time variable.
\begin{figure}[t!]
\centering
\includegraphics[scale=.7]{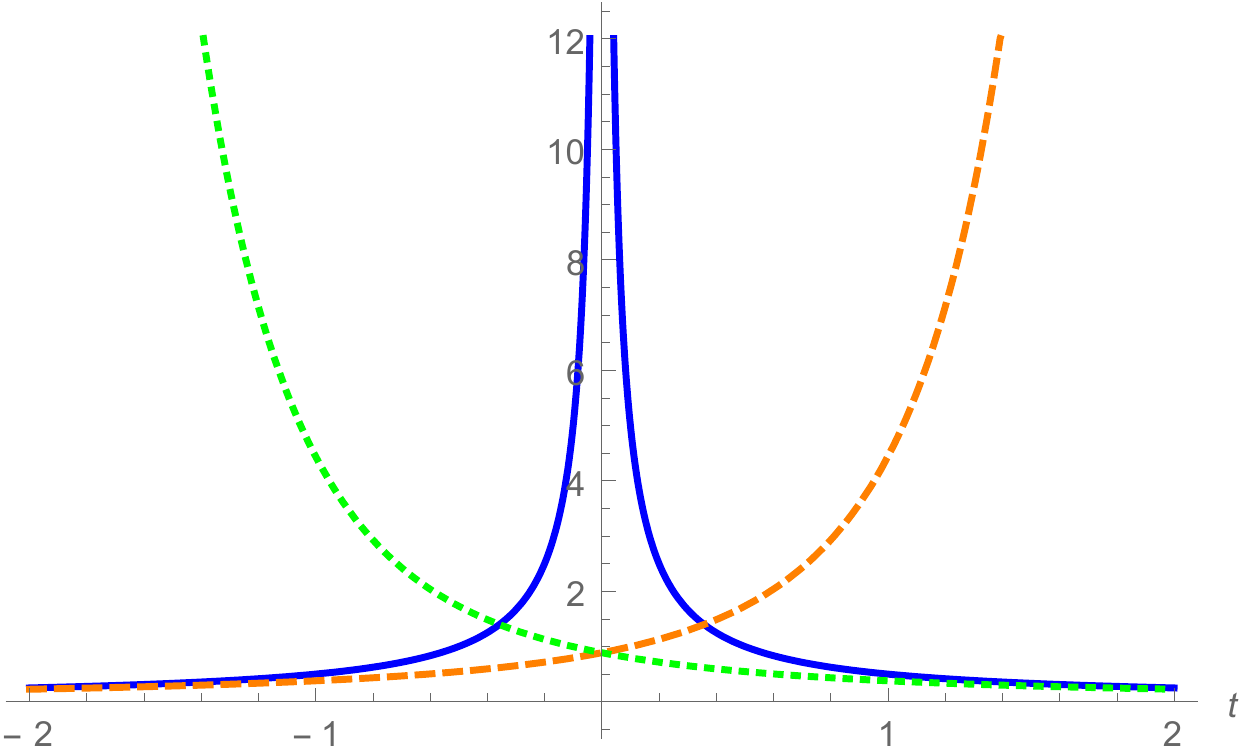}
\caption{Plot of Hubble radius represented by continuous blue curve,
  particle horizon radius in orange dashed curve and event horizon
  radius in green dotted curve for $a=e^{t^2}$ bounce model. Here 
  $a_0=1$ sets the scale-factor at the bounce point and $A$ is assumed
  to have unit magnitude. For more explanation see the text below.}
\label{f:etsq}
\end{figure}

The above expression shows that the particle horizon quickly vanishes
as one goes back in negative time and the particle horizon increases
indefinitely as time evolves, $t \to \infty$.  The plot of the
particle horizon, represented in orange dashed curve, is shown in
Fig.~\ref{f:etsq}. The above result can be interpreted in a simple
way. As the observer moves back in time, $t \to -\infty$, the observer
does not receive any light from the other parts of the universe from
the past as light emitting particles are infinitely distant from the
observer. After a long time the observer first receives light from its
past, and a causal connection is made. At this moment, $t_0$, the
particle horizon $R_P(t_0)$ comes into account. The amount of light
which the observer receives is coming from the closest sources in the
past. As time proceeds more and more regions from the past are coming
in causal contact with the observer and the process continues forever.

In this case we have $|H|=2A|t|$ and consequently one can easily see
that $R_P(t_0)$ is never proportional to $1/|H(t_0)|$.  In the present
case we see that near $t\to -\infty$ both $R_P(t_0)$ and $1/|H(t_0)|$
vanishes. At $t=0$, $R_P(0)$ remains finite whereas $1/|H(t_0)| \to
\infty$. The nature of the Hubble surface, plotted in the blue, is
shown in Fig.~\ref{f:etsq}.  The interesting thing to note is the
relative behavior of the Hubble surface and particle horizon before
the bounce. In the present model the Hubble radius is always greater
than the particle horizon radius before the bounce, unlike the toy
bounce model case, and consequently the particle horizon size
increases with time.  Here $R_H > R_P$ and the two distance scales
differ maximally near the bounce point $t=0$.

We know that the scale-factor changes from the exponential form to
other forms in the expanding phase of the universe. From a purely
mathematical point of view if we assume that the exponential
scale-factor keeps the same functional form as $t\to \infty$ then this
bouncing universe do admit an event horizon. In this case we have
\begin{eqnarray}
R_E(t_0) &=& e^{A t_0^2} \int_{t_0}^{\infty} e^{-A t^2}
dt\nonumber\\ &=& \frac{1}{2}\sqrt{\frac{\pi}{A}}\,\, e^{A t_0^2}
\,\,{\rm erfc}(\sqrt{A}\,\,t_0)\,.
\label{cas1e}
\end{eqnarray}
The behavior of the event horizon is plotted in green in
Fig.~\ref{f:etsq}. The plot shows that the event horizon diverges at
$t_0 \to -\infty$, but in general it has a finite value for all other
times and decreases smoothly as time evolves. As predicted in the last
section, the event horizon steeply decreases during the contracting
phase. In this toy model the event horizon and particle horizon has
the same radius at the bounce point $t=0$.  The nature of the horizons
as plotted in Fig.~\ref{f:etsq} show an interesting feature. As the
bounce is symmetrical in time the particle horizon and the event
horizon are symmetrical in time. The particle horizon transforms into
the event horizon if one changes the direction of time and vice versa.
\section{On the horizon problem in bouncing cosmologies}
\label{hprob}

Inflationary cosmology solves the horizon problem by invoking the
initial inflationary phase during which period the particle horizon
grew exponentially. As a consequence all the regions on the CMBR sky
observed now were in actual causal contact during inflation. Bouncing
models tackle the horizon problem in different ways. One of the ways
to solve the horizon problem in bouncing cosmological models is to
propose that such models may not have any particle horizon, as a
result of which $R_P(t_0) \to \infty$ for any finite time $t_0$. If
the particle horizon does not exist then the observed CMBR sky was in
causal contact, solving the horizon problem of standard Big-Bang
cosmology.  The authors of Refs.~\cite{Battefeld:2014uga,
  Martin:2003bp, Peter:2008qz} strongly favors this idea for solving
the horizon problem in cosmological bouncing scenarios.  In this
scenario the whole of the universe is in principle causally related,
the Hubble radius only limits our observational ability (in the
expanding phase) at any time $t_0$. If the idea regarding the
possibility of the absence of particle horizon in cosmological
bouncing models is elevated to a principle about bouncing models then
we see that our present work shows some counterexamples where the
principle is violated. Our present work shows that some of the
bouncing cosmological models can indeed accommodate particle horizons
for some range of model parameters. If one still wants to keep
$R_P(t_0) \to \infty$ for bouncing models then our work shows the kind
of model parameters which one has to exclude to propose a bouncing
model. A bouncing model may not always satisfy the condition of
nonexistence of the particle horizon for some range of parameters in
the model and consequently one has to avoid the specific parameter
ranges which gives finite $R_P$ to come to a model where $R_P(t_0) \to
\infty$ for all finite values of $t_0$. It must be noted that
any cosmological bounce which starts with a matter dominated phase, as
in Ref.~\cite{Cai:2014bea}, one must have $R_P(t_0) \to \infty$ for
all finite $t_0$ solving the horizon problem in the most straight
forward way. 

In the previous paragraph we have specified the simplest way bouncing
cosmologies can tackle the horizon problem. Unlike the inflationary
paradigm where the horizon problem is solved in all inflationary
models because of the exponential expansion of the particle horizon
(irrespective of model particularities), the bouncing models a priori
do not always give infinite $R_P$. Some models do not have particle
horizon (as LQC model discussed in the article) and some bouncing
models may have particle horizon for some parameter choice (as the
quintom bounce model). If one does not universally accept that all
bouncing models must have $R_P(t_0) \to \infty$ then the horizon
problem acquires new features. If $R_P$ remains always greater than
$R_H$ (except at the bounce point) and $R_P(t_0) \to \infty$ as $t_0
\to -\infty$ in the bouncing universe then the horizon problem can be
solved by assuming that the whole CMBR sky was in causal contact in
the contracting phase of the bouncing universe\footnote{In such cases
  $R_P$ must always be greater than $R_H$ during the contraction phase
  (except very near to $t=0$) was shown in section \ref{pheh}.}. The
whole observable universe at the present moment was in causal contact
long before the bounce solving the horizon problem.  During the
contracting phase the cosmological perturbation modes left the Hubble
horizon and these modes re-entered the Hubble sphere during the
expansion phase producing fluctuations on the CMBR sky. In this case
the Hubble sphere was always causally connected. One can use the above
points to solve the horizon problems in the toy model cases for where
the behavior of the particle horizon are as shown in
Figs.~\ref{f:p2}, \ref{f:hdiff} and \ref{f:quint}.

If the particle horizon distance is bounded in the far past as $t \to
-\infty$, as it happens according to Eq.~(\ref{case1}), then the part
of the universe initially causally connected remains much smaller and
has a finite length scale. In such cases the cosmological
perturbations which will ultimately re-enter the Hubble sphere later
in the expansion phase may produce fluctuations which are not nearly
scale-invariant and more over acausal effects may show up in the
bouncing models\cite{Martin:2003bp}. The acausality appears because
one cannot apply proper initial conditions to all the modes with
various wavelengths at the early phase of contraction. Particularly in
the bouncing case as depicted in Fig.~\ref{f:etsq} although the
particle horizon distance grows and is always more than the Hubble
radius during the expansion phase, the Hubble sphere was not inside
the particle horizon during the contracting phase. Just after the
bounce there are regions inside the Hubble sphere which were never
causally connected before and consequently in this case the horizon
problem is not ideally solved. The cosmological case studied in
subsection \ref{tbm} has other serious difficulties as in this case
the event horizon shrinks in the expanding phase and the Hubble radius
never increases after bounce. The main difficulty in this particular
case is that we have used the same scale-factor throughout all the
phases of the bouncing universe. The scale-factor decreases or
increases very rapidly away from $t=0$ making the cosmological model
pathological.  Ideally bouncing cosmological models should avoid
scenarios where the causal structure is as given in subsection
\ref{tbm} because of the nearly scale-invariant nature of the scalar
perturbations on the CMBR sky and our inherent belief on causality. In
all the toy model examples and the quintom bounce model discussed
previously one can see that $R_P$ diverges as $t_0 \to -\infty$ such
that one can apply initial conditions on all the wavelengths and the
issue of initial scale dependence does not arise.

On the other hand if the particle horizon distance becomes smaller
than the Hubble radius during the expansion phase, as shown in
Fig.~\ref{f:p2} for the $p=2$ case, then deep inside the radiation
dominated phase during expansion (all length scales inside) the Hubble
sphere does not remain causally connected. As long as $R_P(t_0) \to
\infty$ as $t_0 \to -\infty$, as it happens for the above case, the
horizon problem is solved but another important problem arises.  In
this case all the physics related to the formation of CMBR will be
guided by the causal patch predicted by the actual particle horizon
size $R_P$ during the radiation domination. As magnitude of $R_P$
during radiation domination is not defined locally but depends on the
global history of the bouncing model the imprints on CMBR can vary
between various bouncing models. If such a case arises one has to
check properly that the given bouncing model reproduces all the known
features of CMBR spectrum.

The above discussion in this section specifies the various probable
solutions of the horizon problem in the bouncing universe
models. There can be multiple ways one can solve the horizon problem
in such cases. We will now present a general discussion of the
previous results in the next section.
\section{Discussion}
\label{disc}

In the toy cosmological bounce presented in the paper we see that if
the contracting phase prior to the bouncing phase has a power law
scale-factor then the particle horizon exists only when $m>1$. As in
general relativity the exponent in the power law is related with the
barotropic equation of state $\omega$ via
$$\omega=\frac{2-3m}{3m}\,,$$ it is seen that $\omega \ge 0$ only when
$m \le 2/3$, where the equation of state becomes zero at $m=2/3$. If
$m > 2/3$ the barotropic ratio becomes negative. As a consequence it
follows that if there is a power law contraction phase before the
bouncing phase then the condition $\omega<0$ is a necessary condition
for a finite particle horizon. The condition that $m>1$ translates to
\begin{eqnarray}
-1< \omega < -\frac13\,,  
\label{secv}    
\end{eqnarray}  
which specifies that in such a case $\rho + 3P < 0$ during the
contraction phase. This result shows that the particle horizon in such
cases can only exist if the SEC is violated during the power law
contraction phase. This conclusion can also be applied in the quintom
bounce case where the particle horizon can only exist if $r>1/3$ and
consequently $\rho + 3P < 0$ far away from the bouncing point. Near
the bouncing point $\rho + 3P$ becomes a function of time in the
quintom bounce models. At the bounce point one must have to break the NEC to
have a proper bounce in GR based models.

In the toy bounce model we have used various forms of the
scale-factors in the various evolutionary phases of the universe. The
different metric smoothly transforms from one form to the other
because of the junction conditions. The junction conditions and our
choice of the bouncing scale-factor combines to produce an interesting
effect. It is apparent from Eq.~(\ref{tptpp}) that if one chooses
$t^{\prime\prime}=1$, $t^\prime=-1$ and $n=1/2$ then $m$ also turns
out to be $1/2$ when $p$ is an integer. The junction condition and
symmetric matching times combine to predict a symmetrical evolution of
the universe through bounce. If we want to have an asymmetrical
evolution of the universe the matching times should be different which
will lead to dissimilar values of $m$ and $n$ or one may choose to
have $m \ne n$ which will give asymmetric matching times.

In $f(R)$ gravity induced bounce we saw the minima of the particle
horizons appear near $t \to -\infty$. The particle horizons then grows
monotonically. The particle horizon radius displays such a behavior
because the scale-factor at $t \to -\infty$ diverge too fast as one
moves back in time and consequently considerable amount of light
cannot reach any region of the universe. In the toy bounce model and
the quintom bounce model, where particle horizon exists, we observe
that the minima of the particle horizon distance is always near the
bounce point. In these cases the scale-factor during the contraction
phase is given by a power law function of time which diverges as $t
\to -\infty$ but this divergence is much milder than the divergence of
the scale-factor in $f(R)$ gravity induced bounce. As one moves back
in time a much wider part of the universe seems to be causally
connected in the toy bounce model and the quintom bounce model.  But
as in these cases the particle horizon distance exceeds the Hubble
radius the surface defining the particle horizon, at a particular
time, is radially moving inward in a superluminal way. Photons which
are moving radially inward from outside this surface will not be able
to reach this surface.  As the surface contracts with time the
particle horizon distance diminishes with time. It is to be noted that
a contracting particle horizon does not imply that causal regions of
the universe are moving out of the horizon, it implies that as the
universe contracts no new causal regions are entering the particle
horizon.

The minima of the particle horizon appears near the bounce point in
some models, as in the toy bounce model and quintom bounce model. This
happens when $R_P=R_H$. In the example given in this paper the
condition $R_P=R_H$ happens at an instant and consequently the
particle horizon attains a minima at that instant. If the two
surfaces, describing the particle horizon and the Hubble surface,
remain identical for some period of time then during that period the
particle horizon distance will remain constant. This will happen
because light cannot enter radially inward into the particle horizon
as the surface defining the particle horizon, at a particular time,
contracts with the speed of light. More over as the spatial surface
defining the particle horizon, at a particular time, is moving inwards
with just the velocity of light all the emitters inside this surface
move inwards subluminally and the light they emit can reach the
observer in due time. In this case neither the particle horizon
distance increases nor it decreases with time
\footnote{It must be noted that the surface, with definite physical radius,
which define the particle horizon or the Hubble surface at a
particular time, $t$, may not define the particle horizon or the
Hubble surface at a later time $t^\prime>t$. Another surface, with a
different physical radius, will define the particle horizon at
$t^\prime$.  As an example, the surface which coincides with the
particle horizon at time $t$, in the case where $R_P < R_H$, contracts
with time as its physical radial coordinate shrinks with time.  On the
other hand the particle horizon distance increases with time.}.

Although in the Big-Bang paradigm the particle horizon distance and
the Hubble radius are proportional to each other when the scale-factor
of expansion is given by a power law function in bouncing models this
fact does not hold anymore. In the bouncing models the power law
expansion phase may accommodate a particle horizon and the
relationship between the particle horizon distance and the Hubble
radius is much more complex.  Although the present authors have not
seen any article addressing the issue solely related to the
particle horizons in bouncing models, a recent publication
\cite{Barrau:2017ukm} discuss the effect of bouncing models on
luminosity distance in cosmology.

If a bouncing universe does not have a finite particle horizon then
the only compact surface which determines the causal structure of the
universe is the Hubble surface, as in the case of loop quantum
cosmology based bounce studied in this article. Although the Hubble
radius diverges during the bounce time, the Hubble surface is the only
compact 2-dimensional surface which can have any say on the causal
structure of the universe during the contracting and expanding phases.
\section{Conclusion}

The present article addresses the topic related to causality in
general bouncing cosmological models based on the flat FLRW solution.
Keeping the standard definitions of the particle horizon, event
horizon and Hubble radius the present article generalizes their
meaning in a bouncing universe which accommodates an infinitely
stretched (in time) contraction phase.  It is shown that in many
bouncing models the particle horizon may not exist for some parameter
values. When the particle horizon does not exist it means $R_P(t_0)\to
\infty$ for finite $t_0$.In such a case any observer at any time
instant can in principle get light rays from infinite distance away
(in the past) and consequently the horizon problem is solved in a
straight forward manner. If in a bouncing model the particle horizon
does not exist then the Hubble surface defines the causal structure of
this universe. If matter content of the universe, during the
contraction phase, violates some energy energy conditions then the
particle horizon can exist in some simple toy models based on GR.  It
is shown that even if the particle horizon exists one can tackle the
horizon problem in bouncing cosmological models in various cases.

The quintom bounce model shows that under certain circumstances the
quintom universe can have a particle horizon. The casual properties of
the quintom model is similar in nature to toy bounce model. Loop
quantum cosmology induced bounces may not admit any particle
horizons. The bounce model based on $f(R)$ gravity has a single
scale-factor during contraction, bounce and expansion phases. In this
simple model the $f(R)$ gravity based bounce may accommodate the
particle horizon and the event horizon. In reality the $f(R)$ based
model studied in this article will have a particle horizon and the
Hubble radius defined for all time up to the bounce. Depending upon
the future scale-factor (after the bouncing phase) one has to decide
whether the event horizon can exist. In this article we have simply
used the same scale-factor throughout time to show the symmetry
between the particle and event horizons. The example shown in our
paper do not solve the horizon problem and one must reject such cases
while studying bouncing cosmological models.

The toy bounce, which accommodates three phases of evolution of the
universe, exposes the difficulty in calculating the particle horizon
distance as the properties of particle horizon becomes dependent on
the earliest history of the bouncing models.  In the toy model
calculation, presented in the article, the contraction and the
expansion phases are guided by a power law scale-factor where as the
bouncing scale-factor is assumed to be an even function of time. The
simple calculations show that the criterion for existence of particle
horizons depends upon the energy conditions followed by matter during
contraction. We have emphasized the special role of the Hubble surface
which affects the time evolution of the particle horizon.

We have presented a brief discussion on the solution of the horizon
problem in bouncing universes. It is shown than there are number of
ways in which the horizon problem can be solved in bouncing
cosmological models. Some of the probable cosmological models must
have to be modified or abandoned depending upon how successfully they
solve the horizon problem.  The present work shows that the causality
problem in bouncing universe is intrinsically related to an
understanding of the various phases of the universe during the
contraction phase. As our understanding of the contraction phase is
purely speculative at present the models we use to figure out the
nature of particle horizon remains over simplistic. The present
authors believe that although the causality problem in bouncing
universe models are far from being solved the present article shows
the qualitative and quantitative difficulties one must have to
circumvent in the future to produce more meaningful results.

\end{document}